\documentclass[prl,floatfix,showpacs,amsmath,twocolumn]{revtex4}
\usepackage{amssymb}
\usepackage{graphicx}
\usepackage{graphics}

\begin{document}
\input {epsf}

\def\ket#1{|\,#1\,\rangle}
\def\bra#1{\langle\, #1\,|}
\def\braket#1#2{\langle\, #1\,|\,#2\,\rangle}
\def\proj#1#2{\ket{#1}\bra{#2}}

\newcommand{\be}{\begin{eqnarray}}
\newcommand{\ee}{\end{eqnarray}}
\newcommand{\bea}{\begin{eqnarray}}
\newcommand{\eea}{\end{eqnarray}}
\newcommand{\bma}{\begin{subequations}}
\newcommand{\ema}{\end{subequations}}

\title{Quantum memory for non-stationary light fields based on controlled reversible
inhomogeneous broadening}

\author{B. Kraus$^1$, W. Tittel$^1$, N. Gisin$^1$, M. Nilsson$^2$, S. Kr\"oll$^2$, and J.I. Cirac$^3$}

\affiliation{$^1$Group of Applied Physics-Optique, University of Geneva, Switzerland\\
$^2$Division of Atomic Physics, Lund Institute of Technology, Sweden\\
$^3$Max-Planck-Institut f\"{u}r Quantenoptik, Garching, Germany}

\begin{abstract}
We propose a new method for efficient storage and recall of
non-stationary light fields, e.g. single photon time-bin qubits,
in optically dense atomic ensembles. Our approach to quantum
memory is based on controlled, reversible, inhomogeneous
broadening. We briefly discuss experimental realizations of our
proposal.
\end{abstract}
 \maketitle

Photons are good carriers to transmit quantum information, since
they travel fast and only interact weakly with the environment
\cite{reviews}. However, their advantage is also their weakness:
photons can only be used in a probabilistic way for quantum
information processing where interaction between different
carriers is necessary \cite{Knill2001}, and are hard to store.
Therefore, many applications in quantum communication and
information processing call for reversible and efficient mapping
of non-classical photon states onto electronic excitations in
atoms or solids. Such quantum memories enable one to build sources
of single photons on demand based on heralded single photon
sources \cite{Hong1986}, serve as a buffer to store quantum
information, and form a basic ingredient for linear optic quantum
computation \cite{Knill2001}. In addition, quantum memories are
necessary to implement a quantum repeater, which allows for
instance the extension of quantum cryptography to very long
distances \cite{Briegel1998}.

Many different approaches towards quantum memory have been
proposed, based on single absorbers/emitters in a cavity
\cite{single atom quantum memory}, as well as in optically dense
atomic ensembles\cite{ensemble quantum
memory,Moiseev2001,Moiseev2003}. The proposals are either based on
direct transfer of the quantum state of the photon(s) to one of
the atom(s), or "disembodied" transfer via interspecies quantum
teleportation. 
Storage of classical light states has been demonstrated using
photon echoes \cite{photonecho} or based on EIT and halted light
\cite{EIT}, and entanglement between light and atoms has been
reported \cite{light-atom entanglement}. Finally, all-optical
storage of single photons and recall on pseudo-demand has been
shown \cite{Pittman2002}, as well as mapping and storage of a set
of coherent states onto an atomic memory with higher than
classical fidelity \cite{Julsgaard2004}.

The possibility to store and recall single-photon time-bin qubits
at a wavelength of around 1 $\mu m$ would have a large impact on
quantum communication, since they have been shown to be well
suited for transmission in optical fibers over large
distances\cite{reviews,timebin}. Multiple classical light pulses
can be stored and recalled by employing the idea of photon echoes
\cite{photonecho}: The absorption of a spectrally broad pulse in
an inhomogeneously broadened atomic medium leads to a fast
decaying macroscopic dipole moment. Re-phasing and hence
re-emission of the light can be triggered using additional $\pi$
or $\pi /2$ pulses. However, the recall efficiency of photon
echoes normally is only of a few percent
\cite{efficiency,efficiencynote}, preventing this approach from
being used as memory for non-classical light states
\cite{Ohlsson2003}. In 2001, Moiseev and Kr\"oll proposed a novel
scheme, also based on absorption in an inhomogeneously broadened
medium, that enables loss-free storage and reconstruction of a
single-photon wave packet \cite{Moiseev2001}. This proposal
applies to gaseous atomic samples where the inhomogeneous
broadening is given by the Doppler-broadening and takes advantage
of the fact that the Doppler shift is reversed for
counter-propagating pulses. Consequently, the storage time is
limited by the fact that the fast moving atoms eventually escape
from the interaction region with the laser pulses. This time can
be increased by cooling the atoms, however, inhomogeneous
(Doppler) broadening, i.e. the underlying property for storage and
recall on demand will be reduced at the same time. In 2003 Moiseev
\cite{Moiseev2003} extended the idea to sub-millimeter wavelength
photons and impurity atoms in solids, where the storage time is
not limited any more by such "flight-out-of-view" effects. The
protocol takes advantage of spin-lattice interaction and RF pulses
to act on the host atoms and thereby shift the transition
frequencies of the impurities and reverse dephasing.

In this paper we present a new method to efficiently store and
recall non-stationary light fields based on controlled reversible
inhomogeneous broadening (CRIB) in optically dense atomic
ensembles. We demonstrate that by reversing the inhomogeneous
broadening after absorption of an unknown, arbitrary incoming
light state, one can force the atom-light system to evolve
''backwards'' in time. This enables, in principle, unit efficiency
storage and recall of non-classical light states, e.g.
single-photon time-bin qubits. In addition and in contrast to the
two previous proposals \cite{Moiseev2001,Moiseev2003}, we propose
to implement a controlled artificial broadening of an unbroadened
absorption line, based on an interaction with an external field.
This eliminates the difficulty associated with reversing natural,
randomly distributed level shifts which may not always be
possible. Our protocol is general in the sense that it relates to
any interaction that enables controlled, reversible, inhomogeneous
broadening and applies to a variety of materials, e.g. rare earth
ion doped solids, atomic vapor, NV centers, or quantum dots.

In the following, we will describe mathematically the physical
effect on which our scheme is based. We consider a series of
atoms, confined in some region of space, interacting with a pulse
of light propagating along the $+z$ direction, of central
frequency $\omega_0$, and a given polarization. Atom $n$ at
position $z_n$ is initially prepared in a certain ground atomic
level $|g\rangle_n$ and is driven by the laser pulse to another
excited state $|e\rangle_n$ with corresponding transition
frequency $\omega_n$. The Hamiltonian describing this situation
can be written as ($\hbar=1$)
 \bea
 \label{Hamil}
 H&=& \sum_n \omega_n \sigma_n^z + \sum_k c|k| a_k^\dagger a_k
 -\nonumber\\&-& d \sum_n \left[\sigma_n^+ E^+(z_n)+ h.c.\right].
 \eea
Here, the $\sigma$'s are Pauli operators describing the atomic
two--level transition, $a_k$ is the annihilation operator of a
field mode of momentum $k$, and $E^+(z)=i\sum_k g_k e^{ikz} a_k$
is the positive frequency part of the electric field operator. For
the sake of simplicity, we have assumed that the dipole matrix
element $d$ corresponding to the two--level transitions is the
same for all atoms. In (\ref{Hamil}) we have used a 1--dimensional
description, i.e. only included the field modes along the
propagation direction. The rest of the modes are the ones
responsible for spontaneous emission, which will be neglected here
since we assume the total time of the whole considered process to
be much shorter than the spontaneous emission time for each atom.
We further assume that in the incoming electromagnetic field only
the modes with $\omega_k\in [\omega_0-\delta,\omega_0+\delta]$,
with $\delta\ll \omega_0$ are occupied, a fact which has allowed
us to perform the rotating wave approximation in $H$. Accordingly,
we define the slowly varying electric field operators
corresponding to the forward and backward modes as
 \bea
 \epsilon_{f}^+(z,t)&:=& i \sum_{k>0} g_k a_k(t)
 e^{ikz}e^{i(\omega_0 t-k_0 z)},\\
 \epsilon_{b}^+(z,t)&:=& i \sum_{k<0} g_k a_k(t)
 e^{ikz}e^{i(\omega_0 t+k_0 z)},
 \eea
and $\epsilon_{f,b}^-(z,t):=\epsilon_{f,b}^+(z,t)^\dagger$, where
$k_0=\omega_0/c$.

Let us first consider the propagation equation of the light pulse,
which is moving from left to right, through the atomic medium. We
define new atomic operators as
 \begin{subequations}
 \label{atomic1}
 \bea
 s_{n,f}^-(t) &:=& \sigma_n^-(t) e^{i(\omega_0 t-k_0 z_n)}, \\
 s_{n,f}^+(t) &:=& [s_{n,f}^-(t)]^\dagger,\quad s_{n,f}^z(t):=\sigma_n^z(t).
 \eea
 \end{subequations}
and obtain the following Maxwell--Bloch equations \cite{note0}:
 \bea
 \label{MBE}
 i \partial_t s_{n,f}^-(t) &=& - \Delta_n s_{n,f}^-(t) + d s_{n,f}^z(t)
 \epsilon_{f}^+(z_n,t),\nonumber\\
 i \partial_t s_{n,f}^z(t) &=& -d s_{n,f}^+(t)
 \epsilon_{f}^+(z_n,t) + h.c.,\\
 \left[ \partial_z + \frac{1}{c} \partial_t\right]
 \epsilon_{f}^+(z,t) &=& q \sum_{k>0} g_k^2 \sum_n s_{n,f}^-(t)
 e^{i(k-k_0)(z-z_n)},\nonumber
 \eea
where $\Delta_n:= \omega_0-\omega_n$ is the detuning and $q=id/c$.
In the last equation, the sum over $k$ gives rise to a
$\delta(z-z_n)$ which indicates that the atoms are the sources (or
drains) of the electric field. Equations (\ref{MBE}) describe, in
the Heisenberg picture, how the photons in the incoming pulse are
absorbed as they travel through the medium. In particular, if the
medium is optically thick, after a sufficiently long time $t_0$
the photons
will be absorbed. In the Heisenberg picture, this is manifested by
the fact that all expectation values of normally order field
operators will vanish. In the Schr\"odinger picture, the state of
the atom+light system factorizes,
 \be
 \label{Psit}
 |\Psi(t_0)\rangle\simeq |\phi(t_0)\rangle_{\rm atoms}
 \otimes |{\rm vac}\rangle_{\rm field}.
 \ee
Thus, the quantum state of light is stored in the atomic state.
The main issue is now to find a way to recover the state, i.e. to
map it back to the field state. We propose to use the photon--echo
effect, i.e. to instantaneously change some of the atomic
properties such that the field is restored. In order to show how
this works, we just have to look at some symmetry properties of
those equations, and thus we do not need to solve the complicated
Maxwell--Bloch equations nor to make further approximations. The
main idea is to carry out those instantaneous changes so that the
atomic and field operators evolve "backwards in time", i.e. at the
end all the atoms end up in their ground states and the field is
restored but now propagating from right to left.

At time $t_0$, the state of the atoms and field is
$|\Psi(t_0)\rangle$. In the spirit of the proposals in
Ref.\cite{Moiseev2001,Moiseev2003}, we suddenly change the atomic
frequency of each atom from $\omega_n=\omega_0+\Delta_n$ to
$\omega_n=\omega_0-\Delta_n$ (i.e. we change the sign of the
detuning). Simultaneously, a phase shift is applied to each atom
that changes $|g\rangle_n \to e^{2ik_0z_n} |g\rangle_n$. Since we
are going to use the Heisenberg picture for subsequent times, this
is equivalent to keeping the initial state as $|\Psi(t_0)\rangle$
and replacing
 \be
 \label{sigma}
 \sigma_-(t_0)\to \sigma_-(t_0) e^{2ik_0z_n}.
 \ee
Defining the new Heisenberg operators
 \begin{subequations}
 \label{atomic2}
 \bea
 s_{n,b}^-(t) &:=& \sigma_n^-(t) e^{i(\omega_0 t+k_0 z_n)},\\
 s_{n,b}^+(t) &:=& [s_{n,b}^-(t)]^\dagger,\quad
 s_{n,b}^z(t):=\sigma_n^z(t),
 \eea
 \end{subequations}
one can easily obtain the new Maxwell--Bloch equations. Taking
into account that
 \be
 \label{cond}
 \sum_{k>0} g_k^2 e^{i(k-k_0)(z-z_n)} \simeq
 \sum_{k<0} g_k^2 e^{i(k+k_0)(z-z_n)},
 \ee
since, as mentioned before, this expression is basically
proportional to $\delta(z-z_n)$ (i.e., real), we have that the
operators $s_{n,b}^-(t_0+\tau)$, $s_{n,b}^z(t_0+\tau)$ and
$-\epsilon_b^+(z,t_0+\tau)$ fulfill the same differential
equations as $s_{n,f}^-(t_0-\tau)$, $s_{n,f}^z(t_0-\tau)$ and
$\epsilon_f^+(z,t_0-\tau)$, respectively \cite{note}. Let us then
consider a solution for any of the latter in terms of the
operators at time $\tau=0$; for example,
 \be
 \epsilon_f^+(z,t_0-\tau)= \tilde f\left[\tau,s_{n,f}^{\pm,z}(t_0),
 \epsilon_f^{\pm}(z,t_0)\right].
 \ee
We will then have
 \be
 \epsilon_b^+(z,t_0+\tau)= -\tilde f\left[\tau,s_{n,b}^{\pm,z}(t_0),
 -\epsilon_b^{\pm}(z,t_0)\right].
 \ee
Now, we will show that
 \bea
 \label{FEq}
 &&\langle \Psi(t_0)|f\left[\tau,s_{n,b}^{\pm,z}(t_0),
 \epsilon_f^{\pm}(z,t_0)\right]|\Psi(t_0)\rangle =\nonumber\\
 &&\langle \Psi(t_0)|f\left[\tau,s_{n,f}^{\pm,z}(t_0),
 -\epsilon_b^{\pm}(z,t_0)\right]|\Psi(t_0)\rangle
 \eea
for all (analytic) functions $f$, and therefore that the
expectation values of any observable at time $\tau$ will give the
same result. This implies the desired result, namely that the
evolution of the system at time $t_0+\tau$ will be closely
connected to that at $t_0-\tau$; in particular, at time $2t_0$ we
will recover the initial pulse propagating in the backward
direction and with a global $\pi$ phase shift. In order to prove
Eq. (\ref{FEq}) we notice first that
$s_{n,b}^{\pm,z}(t_0)=s_{n,f}^{\pm,z}(t_0)$, as it can be seen
from (\ref{atomic1}) and (\ref{atomic2}) and taking into account
(\ref{sigma}). On the other hand, if we expand $f$ in the left
hand side of Eq. (\ref{FEq}) in normally ordered powers of
$\epsilon_f^{\pm}(z,t_0)$ all the terms except the zeroth power
will vanish as they are evaluated in the vacuum $|{\rm
vac}\rangle$ [c.f. (\ref{Psit})]. These zeroth order appear when
we commute $\epsilon_f^{-}(z,t_0)$ with $\epsilon_f^{+}(z',t_0)$.
Analogously, in the right hand side of Eq. (\ref{FEq}) only the
zeroth order terms will survive, and they will appear when we
commute $\epsilon_b^{-}(z,t_0)$ with $\epsilon_b^{+}(z',t_0)$.
However, using their definitions one can readily see that
 \be
 \left[\epsilon_f^{-}(z,t_0),\epsilon_f^{+}(z',t_0)\right]
 \simeq \left[\epsilon_b^{-}(z,t_0),\epsilon_b^{+}(z',t_0)\right]
 \ee
as it follows from (\ref{cond}). Note also that this explanation
becomes much simpler in the semiclassical case, where the electric
field operators are substituted by c--numbers; in that case,
$\epsilon_f^+(z,t_0-\varepsilon)=-\epsilon_b^+(z,t_0+\varepsilon)\to
0$ as $\varepsilon\to 0$, since at time $t_0$ the field has been
completely absorbed, and thus
$\epsilon_b^+(z,t_0+\tau)=-\epsilon_f^+(z,t_0-\tau)$.

Thus, according to this analysis, one should proceed as follows:
first, one should prepare a single atomic absorption line and then
broaden it inhomogenously up to a value of $\delta$. This has to
be carried out in a controlled way such that later on one can
change the sign of the detunings of all the atoms. Second, the
light pulse of duration $\delta t_{light}$ is sent into the
optically thick atomic ensemble and is completely absorbed
\cite{optical depth}.
In order to retrieve the pulse the position dependent phase--shift
is imprinted in all atoms, and the sign of the detunings are
suddenly changed during a time $\delta t$. The whole process
should happen within a time, much shorter than the atomic
spontaneous emission time 
or any other decoherence time. Thus, the conditions for this
scheme to work are $\delta t\ll \delta t_{light}, t_0\ll
T_{decoh}$. Note that after absorption and sufficient
(inhomogeneous) dephasing, one might also remove the inhomogeneous
broadening and reestablish it with opposite sign when readout is
desired. Note further that in order to extend the storage time
beyond $T_{decoh}$, one can coherently transfer the excited state
population to another internal atomic ground state with extended
coherence time until the time one wants to retrieve the pulse, in
which this operation is reversed. Moreover, for
counter-propagating beams, this process can automatically lead to
the position dependent phase--shift \cite{pi pulses}.

In the following, we will briefly discuss several experimental
realizations of our proposal. As mentioned before, CRIB requires
an atomic medium with a short decoherence time, i.e. a small
homogeneous linewidth. For instance, alkaline metals with typical
linewidths in the MHz range enable storage during a few hundred
ns, and linewidths in rare-earth ion doped (RE) crystals at
cryogenic temperatures may be below 100 Hz, i.e. storage times of
several ms have been reported \cite{linewidths,Fraval2004}.

First, an isolated absorption line on a non-absorbing background
has to be prepared. To this aim, the natural inhomogeneous
broadening has to be suppressed. Regarding free atoms, the Doppler
broadening can be decreased by cooling. In the case of RE doped
crystals, inhomogeneous broadening is caused by the fact that the
ions are located in slightly different surroundings in the host.
Through optical pumping, it is possible to transfer ions absorbing
at ''undesired'' frequencies to other long-lived (hyperfine)
levels and thus to empty a spectral region of absorbing ions and
to form the desired narrow absorptive feature
\cite{opticalpumping}.

Next, the absorption line has to be broadened in a controlled and
reversible way. Towards this end, one can take advantage of the
interaction between atoms and a magnetic or an electric field,
i.e. the Zeeman or Stark shift. Note that the maximum broadening
has to remain smaller than the hyperfine 
or fine structure splitting, respectively. Regarding the Zeeman
effect, typical level shifts are of the order of the Bohr magneton
divided by Planck's constant, i.e. $\approx 13$ MHz/mT. When using
atoms (ions) featuring a permanent dipole moment, one might also
take advantage of a dc electric field. For instance, in the case
of RE doped crystals, shifts of the order of $100$ kHz/(Vcm-1)
have been observed \cite{dcshift}. Application of a magnetic, or
electric (dc) field gradient leads thus to a desired position
dependent detuning, $\Delta$, that can be reversed by inverting
the field \cite{2kz}. More details regarding CRIB based on
dc-Stark shifts in RE doped crystals can be found in
\cite{Nilsson2004b}. To complete these examples, the energy levels
of neutral atoms (lacking a permanent dipole moment) can be
shifted in a controlled way employing ac electric fields, e.g.
strongly detuned light fields (ac-Stark-- or light shift). In this
case, the shift depends linearly on the light intensity and is
inversely proportional to the detuning of the light field with
respect to the unperturbed absorption line. Typical line shifts,
for instance for the Caesium D1 line, are of the order of 200
MHz/(10$^9$Wm$^{-2}$) at a laser detuning of 10 nm. The shifts can
be reversed by changing the detuning of the laser to its inverse
value.





In this paper, we proposed a new protocol for quantum state
storage for non-stationary light fields in an inhomogeneously
broadened atomic ensemble. Instead of solving the underlying
equations of motion, we showed that by applying suitable phase and
frequency shifts once the light field has been absorbed, the
atom-light system will evolve backwards in time. In order to
implement these shifts, the inhomogeneous broadening must be
controllable and reversible. We briefly discussed several
experimental realizations of CRIB. To conclude, let us mention
that it is possible to acquire information about successful
loading of the memory by employing teleportation-based state
transfer, along the lines exploited in quantum relays
\cite{reviews,relay}. This is an important necessity to implement
a quantum repeater, and also enables to deterministically entangle
distant atomic ensembles for quantum networks by starting with
entangled photon pairs. Note also that CRIB enables storage of a
whole sequence of light states and recall in inverse order. In
addition, using several broadened, well separated absorption
lines, frequency multiplexing can be employed. This enables recall
of photonic qubits in arbitrary order by triggering only rephasing
of atoms within a desired range of resonance frequencies. Finally,
it is likely that tailored rephasing can also be used to implement
general "interatomic" unitary transformations \cite{Bai84} or
measurements, e.g. pulse compression \cite{pulse compression}.

Financial support by the Swiss NCCR Quantum Photonics, the Swedish
Research Council, and by the EU IST-FET ESQUIRE, RESQ and
COVAQUIAL projects is acknowledged.


\begin{references}
\bibitem{note0} In these equations we have omitted the backward field
$\epsilon_b(z,t)$, i.e. we have ignored back--scattering, which
can be shown to be negligible.
\bibitem{note} The forward terms $-\epsilon_f^+(z,t_0+t)$ can be
neglected using the same arguments that allowed us to neglect the
backward scattering terms (c.f. \cite{note0}) in Eq. (\ref{MBE}).
\bibitem{optical depth}Note that the optical depth can be
        increased by taking advantage of cavity enhanced light-atoms
        interaction, however, without the constraint of strong coupling as
        necessary in single atom based schemes \cite{single atom quantum memory}.
\bibitem{2kz}A linear, z-dependent detuning $\Delta(z)$, can also be used to achieve the position
dependent phase shift by extending the atom-field interaction
beyond absorption by a time $\delta\tau$ with
$\Delta(z)\delta\tau=2kz$.
\end{references}
\end{document}